# Evaluation Study for Delay and Link Utilization with the New-Additive Increase Multiplicative Decrease Congestion Avoidance and Control Algorithm

**HAYDER NATIQ JASEM, ZURIATI AHMAD ZUKARNAIN, MOHAMED OTHMAN, SHAMALA SUBRAMANIAM**
(Department of Communication Technology and Networks
Faculty of Computer Science and Information Technology
University Putra Malaysia, 43400 UPM, Serdang, Selangor, Malaysia
E-mail: hayder_n@yahoo.com)

**Abstract:** As the Internet becomes increasingly heterogeneous, the issue of congestion avoidance and control becomes ever more important. And the queue length, end-to-end delays and link utilization is some of the important things in term of congestion avoidance and control mechanisms. In this work we continue to study the performances of the New-AIMD (Additive Increase Multiplicative Decrease) mechanism as one of the core protocols for TCP congestion avoidance and control algorithm, we want to evaluate the effect of using the AIMD algorithm after developing it to find a new approach, as we called it the New-AIMD algorithm to measure the Queue length, delay and bottleneck link utilization, and use the NCTUns simulator to get the results after make the modification for the mechanism. And we will use the Droptail mechanism as the active queue management mechanism (AQM) in the bottleneck router. After implementation of our new approach with different number of flows, we expect the delay will less when we measure the delay dependent on the throughput for all the system, and also we expect to get end-to-end delay less. And we will measure the second type of delay a (queuing delay), as we shown in the figure 1 bellow. Also we will measure the bottleneck link utilization, and we expect to get high utilization for bottleneck link with using this mechanism, and avoid the collisions in the link.

## 1 Introduction

End-to-end congestion avoidance and control as well as fair network resource management would have had great benefit had the TCP sender known of the behavior of the bottleneck queue and the delay in this queue.
Several methodologies have been developed to estimate bandwidth and bottleneck queue based on temporary measurements of throughput, inter-packet gap, or RTT. For example, TFRC [Handley, 03] calculates throughput via a throughput equation that incorporates the loss event rate, round-trip time and packet size. TCP-Vegas [Brakmo, 95] estimated the level of congestion using throughput-based measurements.
TCP-Vegas demonstrated that measurement-based window adjustments is a viable mechanism, however, the corresponding estimators can be improved. In TCP-Westwood [Casetti, 02], the sender continuously measures the effective bandwidth used by monitoring the rate of returned ACKs. TCP-Real [Tsaoussidis, 02] uses wave patterns: a wave consists of a number of fixed-sized data segments sent back-to-back, matching the inherent characteristic of TCP to send packets back-to-back. The protocol computes the data-receiving rate of a wave, which reflects the level of contention at the bottleneck link. Bimodal congestion avoidance and control mechanisms [Attie, 03] compute the fair-share of the total bandwidth that should be allocated for each flow at any point during the system's execution.
Additive Increase/Multiplicative Decrease (AIMD) is the algorithm that controls congestion in the Internet [Chiu, 89]. It is coded into TCP and adjusts its sending rate mechanically, according to the 'signals' TCP gets from the network.
AIMD-based congestion avoidance and controls [Lahanas, 03] developed the AIMD algorithm to AIMD-FC to get more efficiency and fairness than the AIMD algorithm. TCP-Jersey [Xu, 04] operates based on an "available bandwidth" estimator to optimize the window size when network congestion is detected. The Packet-Pair technique [Keshav, 91] estimates the end-to-end capacity of a path using the difference in the arrival times of two packets of the same size traveling from the same source to the same destination. The TCP-based New-AIMD congestion avoidance and control [Hayder, 08] developed the AIMD algorithm into the New-AIMD to get more efficiency and fairness than the AIMD-FC+ algorithm and evaluated the efficiency compared to AIMD-FC+ in [Lahanas, 03], [Lahanas, 02]. In [Hayder, 09] they investigated the fairness of New-AIMD and evaluated it compared to AIMD-FC+ [Lahanas, 03]. And now in this work we want to investigate and evaluate the implementations of the New-AIMD algorithm in TCP on the network to avoid and control any congestion, and to keep the queue size less than the queue size in the related work to reduce the delay for data transmission in the network system, And also to get high utilization for bottleneck link.

# 2 Congestion Control

It was not until 1988 that a widely accepted congestion control algorithm was finally suggested [Jacobson, 88]. This algorithm employed the Additive Increase Multiplicative Decrease (AIMD) principle. According to the AIMD, a protocol should increase its sending rate by a constant amount and decrease it by a fraction of its original value, each time an adjustment is necessary. This mechanism is the base of virtually all TCP implementations used in the Internet today, since it is proven to converge on both a desirable level of efficiency as well as a desirable level of fairness among competing flows [Chiu, 89].
In the years that followed the establishment of AIMD as the standard algorithm to be used in TCP, the Internet underwent numerous changes and rapidly increasing popularity. With the availability of widespread services such as e-mail and the World Wide Web (WWW), the Internet became accessible to a broader range of people, including users lacking any particular familiarity with computers. Although new competing technologies emerged and the demands from a transport layer protocol were highly increased, TCP not only survived but also became an integral ingredient of the Internet, experiencing only minor modifications. These modifications reflect the different in-use TCP versions (TCP-Tahoe, TCP-Reno, TCP-NewReno) [Jacobson, 88], [Allman, 99], [Floyd, 99], experimental TCP versions (TCP-SACK, TCP-Vegas) [Mathis, 96], [Brakmo, 95], as well as special-purpose TCP versions (T/TCP) [Braden, 94].

## 2.1 The AIMD Principle

As mentioned earlier, the basic concept of AIMD was proven to yield satisfactory results when the network infrastructure consisted of hard-wire connected components. One year after the appearance of AIMD in 1988, the authors in [Chiu, 89] provided a detailed analysis of different congestion control strategies, as well as what makes the existence of such a strategy in a transport protocol crucial. Below we give a few important points made in this work.
The major issue of concern to a transport protocol is its efficiency. On a network link crossed by a number of different flows running the same protocol, the ideal situation is to utilize as much of the available bandwidth without introducing congestion (i.e. packets queuing up on the router). In Fig. 1, we see the achieved throughput as a function of the network load. It becomes clear that we need to avoid overloading the link, since the achieved throughput will diminish. For a protocol to operate in the area between the points labeled as Knee and Cliff, a congestion control mechanism is necessary. In [Chiu, 89] efficiency is defined as the closeness of the total load to the Knee, which is a good starting point.

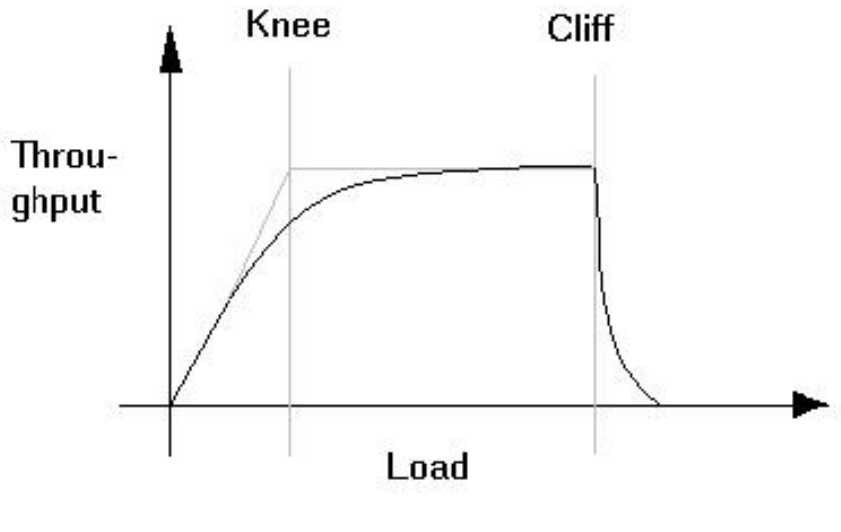

*Fig. 1: Throughput as a function of load.*

Besides utilizing a high portion of the available bandwidth, a transport protocol must also be fair to the rest of the flows traversing the same part of the network. An efficient transport protocol does not necessarily mean that it is also fair. A single flow might take up the largest portion of the available bandwidth while the rest remain idle. Obviously, this is an undesirable behavior and in certain cases, gaining higher fairness is worthwhile even at the cost of reduced efficiency.
Intuitively, fairness is the closeness of the throughput achieved by each flow to its fair share.

### 2.2 System Model

Chiu and Jain [Chiu, 89] formulated the congestion avoidance problem as a resource management problem and proposed a distributed congestion avoidance mechanism named 'additive increase/multiplicative decrease' (AIMD). In their work, as a network model, they used a "binary feedback" scheme with one bottleneck router [Ramakrishnan, 90], as shown in Figure2. It consists of a set of *m* users, each of which sends data in the network at a rate [2] $w_i$. The data send by each user are aggregated in a single bottleneck and the network checks whether the total amount of data send by users exceeds some network or bandwidth threshold $X_{goal}$ (we can assume that $X_{goal}$ is a value between the knee and the cliff and is a characteristic of the network). The system sends a binary feedback to each user telling whether the flows exceed the network threshold. The system response is 1 when bandwidth is available and 0 when bandwidth is exhausted.
The feedback sent by the network arrives at the same time for all users. The signal is the same for all users and they take the same action when the signal arrives. The next signal is not sent until the users have responded to the previous signal. Such a system is called a synchronous feedback system, or simply a synchronous system. The time elapsed between the arrival of two consecutive signals is discrete and the same after every signal arrival. This time is referred to as RTT.
The system behavior can be defined the following time units:
A step (or round-trip time – RTT) is the time elapsed between the arrival of two consecutive signals.
A cycle or epoch is the time elapsed between two consecutive congestion events (i.e., the time immediately after a system response 0 and ending at the next event of congestion when the system response is again 0).

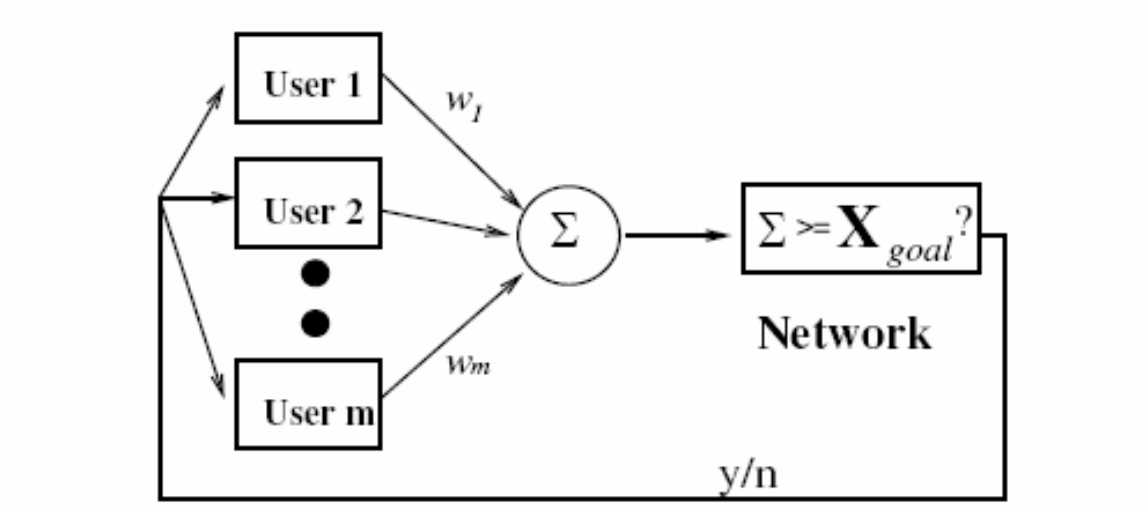

*Fig. 2: A control system model of m users sharing a network.*

In practice, the parameter $X_{goal}$ is the network capacity (i.e. the number of packets that the link and the router buffer can hold – or on-the-fly packets). When the aggregate flow rate exceeds the network capacity the flows start to lose packets. If the transport protocol provides reliability mechanisms (e.g. as in TCP) it can detect the packet loss or congestion event. Since the majority of the applications use reliable transport protocols (e.g. TCP), the binary feedback mechanism has an implicit presence; a successful data transmission is interpreted as available bandwidth, and a packet loss is interpreted as a congestion event [Jacobson, 88].
Algorithmically the AIMD can be expressed with the following lines:

```
AIMD ()
α_i : constant = packet-size()
W : integer // congestion window
repeat forever
        send W bytes in the network
        receive ACKs
```

if W bytes are ACKed

W ← W + $\alpha_i$

else

W ← W/2

end

END-AIMD

**2.3 A pseudocode of New-AIMD**

Let us assume network capacity (Window size or $X_{goal}$) is *W*. For Simplicity let us assume we have two flows system f1 and f2. Initially let flows f1 and f2 contain $x_1$ and $x_2$ window respectively. With out loss of generality we assume that $x_1 < x_2$ and $x_1 + x_2 <$ W furthermore, we are assuming that system converges to 'fair' in 'm' cycle. In 1st cycle Pseudocode is given by total flow is:

$x_1 + x_2 + 2k_1$ (1)

In AIMD is $x_1 + x_2 + 2k_1$

It is clear in 1st cycle that system has $k_1$+1 Round Trip Time (RTTs) or steps. Let $x_1 + x_2 + 2k_1 \geq W$ then there is Congestion and system gives 0 feedback. Now we will use decrease step. In 2nd cycle Pseudocode is given by total flow is:

$\frac{x_1}{2} + \frac{x_2}{2} + 2k_1 + 2k_2$ (2)

In AIMD is $\frac{x_1}{2} + \frac{x_2}{2} + k_1 + 2k_2$

Obviously 2nd cycle contains $k_2 + 1$ RTT. Let $\frac{x_1}{2} + \frac{x_2}{2} + 2k_1 + 2k_2 \geq W$ then system gives 0 feedback. Obviously we will use decrease step. In 3rd cycle Pseudocode is given by total flow is:

$\frac{x_1}{2^2} + \frac{x_2}{2^2} + 2k_1 + 2k_2 + 2k_3$ (3)

In AIMD is $\frac{x_1}{2^2} + \frac{x_2}{2^2} + k_1 + k_2 + 2k_3$

Here 3rd cycle contains $k_3 + 1$ RTTs. Let $\frac{x_1}{2^2} + \frac{x_2}{2^2} + 2k_1 + 2k_2 + 2k_3 \geq W$ then system gives 0 feedback. Obviously we will use decrease step. Similarly at $m^{th}$ cycle we have total flow is:

$\frac{x_1}{2^{m-1}} + \frac{x_2}{2^{m-1}} + 2k_1 + 2k_2 ... 2k_m$ (4)

In AIMD is $\frac{x_1}{2^{m-1}} + \frac{x_2}{2^{m-1}} + k_1 + k_2 ... 2k_m$

Suppose $m^{th}$ cycle points to equilibrium that is all flows share fair allocation of resources.

The algorithmic approach when initial window size of 2 flows and Window size are $x_1, x_2, W$ respectively, is given by:

AIMD ( $x_1, x_2, W$ )

$z = x_1 + x_2$ // $z$ denotes used Capacity of Network.

$k = 1, t = 1$ // $k$ denotes numbers of RTTs

while (1)

{

k = k + 1

$z = x_1 + x_2 + 2t$
t = t + 1
if (z >= W)
{
$x_1 = \frac{x_1}{2}$
$x_2 = \frac{x_2}{2}$
$z = x_1 + x_2 + 2t$
k = k + 1
}
}
END-AIMD

Total number of packets in various cycles:
In 1st cycle, total number of packets is given by: $(k_1 + 1)(x_1 + x_2 + 2k_1)$, But from 1st cycle we have $x_1 + x_2 + 2k_1 = W$ Therefore $x_1 + x_2 + k_1 = W - k_1$.
Thus total number of packet is given by $(1 + k_1)(W - k_1)$.
In 2nd cycle, total number of packets is given by: $(1 + k_2)(\frac{x_1}{2} + \frac{x_2}{2} + 2k_1 + k_2)$.
But from 2nd cycle we have: $(\frac{x_1}{2} + \frac{x_2}{2} + 2k_1 + 2k_2) = W$.
Therefore $\frac{x_1}{2} + \frac{x_2}{2} + 2k_1 + k_2 = W - k_2$.
Thus total number of packets is given by: $(1 + k_2)(W - k_2)$.
Similarly in 3rd cycle, total number of packets is given by: $(1 + k_3)(W - k_3)$.
Similarly in $m^{th}$ cycle, total number of packets is given by: $(1 + k_m)(W - k_m)$.
Thus total number of packets in all cycles is given by: $(1 + k_1)(W - k_1)$ + $(1 + k_2)(W - k_2)$ + $(1 + k_3)(W - k_3)$ + …+$(1 + k_m)(W - k_m)$.
But from equation 1 we have: $k_2 = (W - 2k_1)/4$.
From equations 2 and 3 we have: $k_3 = \frac{1}{2}k_2$.
From equation 3 and 4 we have: $k_4 = \frac{1}{2}k_3$, $k_4 = (\frac{1}{2^2})k_2$ Thus $k_m = (\frac{1}{2^{m-2}})k_2$ for m=3.
The Additive Increase/Multiplicative Decrease (AIMD) and (New-AIMD) algorithms are described by details in [Lahanas, 03], [Hayder, 08], [Hayder, 09].

## 3 Delays

**Delay** and **latency** are similar terms that refer to the amount of time it takes a bit to be transmitted from source to destination [Michael, 05].
**Jitter** is delay that varies over time. One way to view latency is how long a system holds on to a packet. That system may be a single device like a router, or a complete communication system including routers and links [Ravi, 08], [Michael, 05].

Closely related topics include bandwidth and throughput. These are illustrated in Figure 3. Bandwidth is often used to refer to the data rate of a system, but it appropriately refers to the width of the frequency band that a system operates in. Data rate and wire speed are better terms when talking about transmitting digital information. The speed of the system is affected by congestion and delays. Throughput refers to the actual measured performance of a system when delay is considered.

Delays are caused by distance, errors and error recovery, congestion, the processing capabilities of systems involved in the transmission, and other factors.

Delay of distance (called propagation delays) is especially critical when transmitting data to farther destinations. Most communications require a round-trip exchange of data, especially if the sender is waiting for an acknowledgement of receipt from the receiver. Increasing data rates allows you to send more bits in the same amount of time, but it doesn't help improve delay. Excessive delay may cause a receiving system to time out and request a retransmission. The delay factor has to be adjusted when excessive delay exists [Michael, 05].

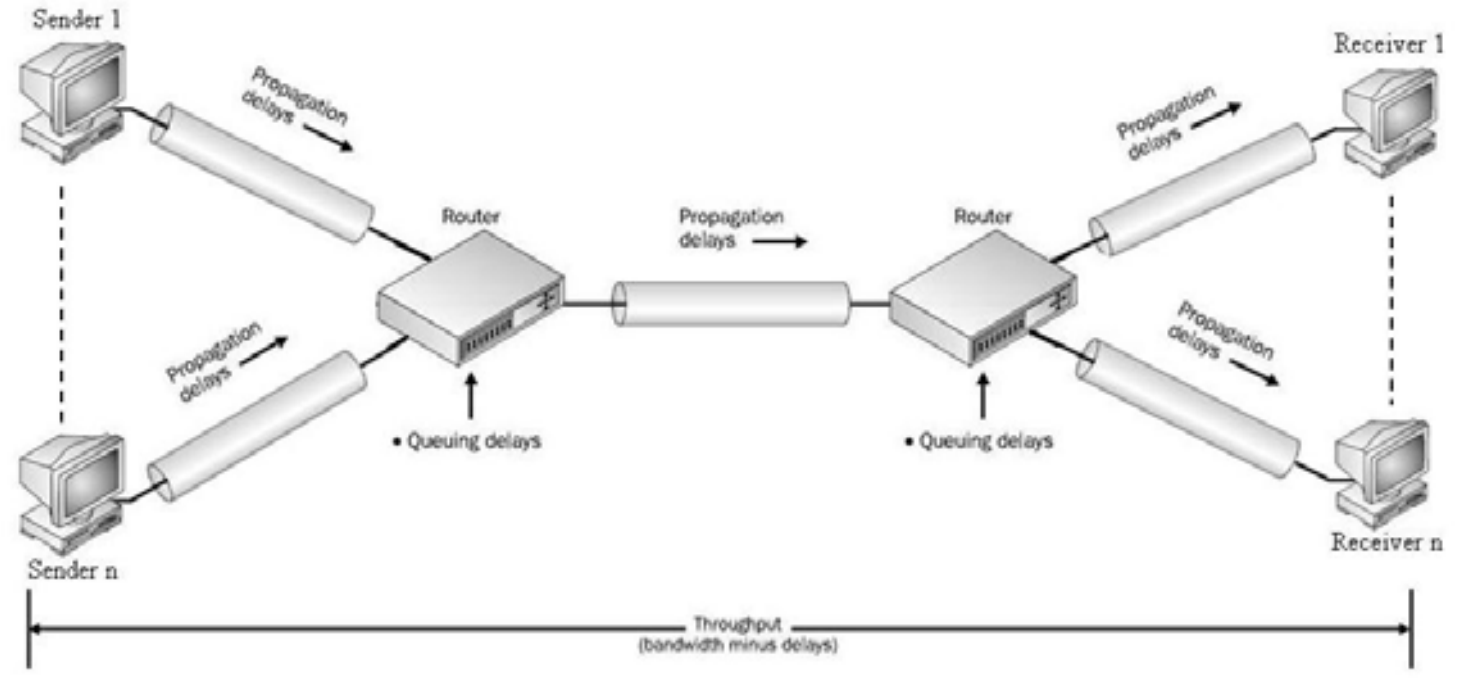

*Fig. 3: The relationships between bandwidth, delay, and throughput*

Delay is problematic for real-time traffic like interactive voice calls and live video. Delay can also be a problem with time-sensitive transaction processing systems. Delay caused by congestion must be avoided, so bandwidth management, priority queuing, and QoS are important to ensure that enough packets get through on time.

Variation in delay (jitter) is more disruptive to a voice call than the delay itself [Johari, 01], [Brakmo, 95].

### 3.1 Causes of Delay

Delay is caused by hardware and software inefficiencies, as well as network congestion and transmission problems that cause errors. Delay may be caused by the following:

- Network congestion, caused by excessive traffic.
- Processing delays, caused by inefficient hardware.
- Queuing delays occur when buffers in network devices are flooded.
- Propagation delay is related to how long it takes a signal to travel across a physical medium [Michael, 05].

In this work we will investigate and focus on the network congestion delay and queuing delay from different causes of delays.

### 3.2 Congestion Delay

As traffic increases on the network, congestion increases. Congestion occurs at routers and switches, causing delay that is variable (jitter).

Ethernet shared medium is prone to congestion. A user must wait if the cable is being used, and collisions occur if two people try transmitting at the same time. Both users wait and then try again, causing further delay for the end-user application (end-to-end delay) [Jahwan, 03].

When a TCP/IP host begins to transmit, it has no way to monitor the network for downstream congestion problems. The host cannot immediately detect that a router is becoming overburdened.
Only when the sender is forced into retransmitting dropped packets does it get a sense that the network must be busy and then start to slow down its transmissions.
Several techniques [Michael, 05] have been developed to resolve congestion problems on TCP/IP networks, such as slow start and congestion avoidance. Congestion controls help hosts adapt to traffic conditions. A transmission starts slowly and builds up until congestion is detected.

### 3.3 Queuing Delay

After a router receives and examines a packet, it sends the packet to a buffer where it is queued up for transmission, usually on a first-in, first-out (FIFO) basis. Routers receive packets from many different sources, so the devices can easily be overwhelmed.
Buffers start to fill up when the network gets busy. Traffic may move into a queue faster than it can be moved out. If packets are delayed long enough, the source systems may begin retransmitting packets under the assumption that packets have been lost. This adds to network congestion and delay [Lin, 05], [Wang, 06].
As mentioned, queues are usually processed on a first-come, first-served basis. Priority queuing techniques give some packets precedence over others. Packets may be marked or tagged in advance so that they are directed into a queue that matches their priority. Alternatively, a device may examine packet contents to determine priority [Michael, 05] [Wang, 06], [Eitan, 04].

## 4 Network Utilization

Network utilization is the ratio of current network traffic to the maximum traffic that the port can handle. It indicates the bandwidth use in the network. While high network utilization indicates the network is busy, low network utilization indicates the network is idle. When network utilization exceeds the threshold under normal condition, it will cause low transmission speed, intermittence, request delay and so on.
Networks of different types or in different topology have different theoretical peek value under general conditions. However, this doesn't mean that the higher network utilization is the better. We must make sure there is no packet loss when network utilization reaches a certain value. For a switched Ethernet, 50% network utilization can be considered as high efficiency. If using router or hub as core switch device in the network, the network utilization should be lower than the link bandwidth capacity to avoid the increasing collisions. Through monitoring network utilization, we can understand whether the network is idle, normal or busy [Ravi, 08], [Michael, 05].
And also, a file of size $f$ with a total transfer time of $\Delta$ on a TCP connection results in a TCP transfer throughput denoted by r and obtained from equation (5)

$$r = f / \Delta \qquad (5)$$

We can also derive the bandwidth utilization, $\rho u$, assuming that the link bandwidth is B, by equation (6)

$$\rho u = r / B \qquad (6)$$

In our approach when we implement the New-AIMD mechanism we try to get high bottleneck link utilization for link capacity (network resources). And we will make many experiments depends on number of flows using the link at same time, to show the different between them.

## 5 Drop Tail AQM Algorithm

Drop Tail (DT) is the simplest and most commonly used algorithm in the current Internet gateways, which drops packets from the tail of the full queue buffer. Its main advantages are simplicity, suitability to heterogeneity and its decentralized nature.
However this approach has some serious disadvantages, such as no protection against the misbehaving or non-responsive flows (i.e., flows which do not reduce their sending rate after receiving the congestion signals from gateway routers) and no relative Quality of Service (QoS).
The QoS is idea in the traditional "best effort" Internet, in which we have some guarantees of transmission rates, error rates and other characteristics in advance. QoS is of particular concern for the continuous transmission of high-bandwidth video and multimedia information. Transmitting this kind of content is difficult on the present Internet with DT.
Generally DT is used as a baseline case for assessing the performance of all the newly proposed gateway algorithms [Aun, 03], [Eitan, 04].

# 6 National Chiao Tung University Network Simulator

The NCTU network simulator is a high-fidelity and extensible network simulator and emulator capable of simulating various protocols used in both wired and wireless IP networks. The NCTUns can be used as an emulator, it directly uses the Linux TCP/IP protocol stack to generate high-fidelity simulation results, and it has many other interesting qualities. It can simulate various networking devices. For example, Ethernet hubs, switches, routers, hosts, IEEE 802.11 wireless stations and access points, WAN (for purposely delaying/dropping/reordering packets), optical circuit switch, optical burst switch, QoS DiffServ interior and boundary routers. It can simulate various protocols for example, IEEE 802.3 CSMA/CD MAC, IEEE 802.11 (b) CSMA/CA MAC, learning bridge protocol, spanning tree protocol, IP, mobile IP, Diffserv (QoS), RIP, OSPF, UDP, TCP, RTP/RTCP/SDP, HTTP, FTP and telnet [Wang, 07].

# 7 Experimental Methodology

We have implemented our evaluation plan on the NCTUns network simulator. The network topology used as a test-bed is the typical single-bottleneck dumbbell, as shown in Figure 4.
For the simulation scenario as a general case we will have the following setup details:
The link's capacity at the senders, receivers and bottleneck link is 5Mbps. We used an equal number of senders and receivers nodes. All DT queues have 100-packet lengths. The link distance between nodes is 3000 meter. And we will use the TCP-SACK with New-AIMD to evaluate the algorithm performance.

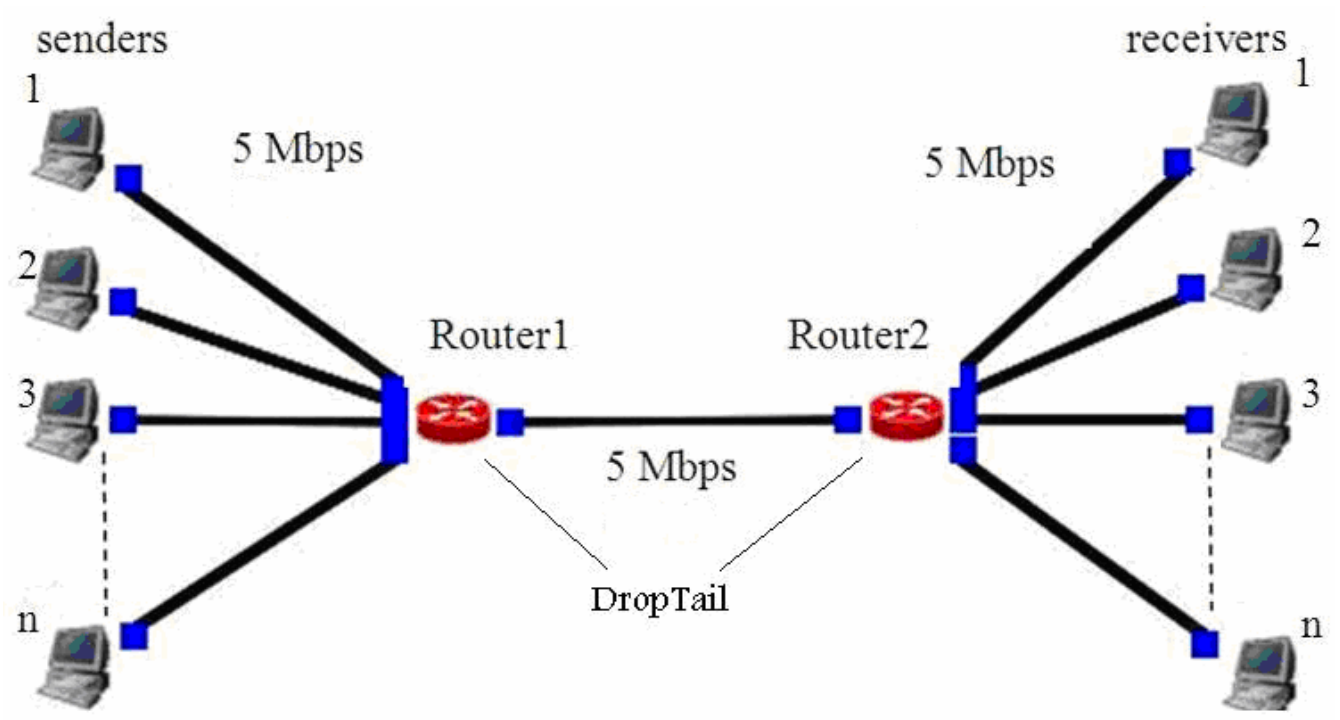

*Fig. 4: Multiple flow experimental set-up for New-AIMD evaluation*

# 8 Results of Simulation

## 8.1 The First Experiment: Results

In the following figures (Fig. 3, 4), we supposed the maximum data size that we want to transmit it from the sender to the receiver is equal to 20000 KB. After complete data transmit to the receiver we can calculate the total time that takes to do the data transmission, as we mentioned above about the relation between the throughput and delay. Also, we have different cases for calculating the time, and the results will depend on the number of flows in the bottleneck (1, 2, 3, 4, or 5 flows) at the same time. Also in the figure 4 we shown the comparison between our mechanism New-AIMD and AIMD-FC+ in the previous related work [Lahanas, 03].

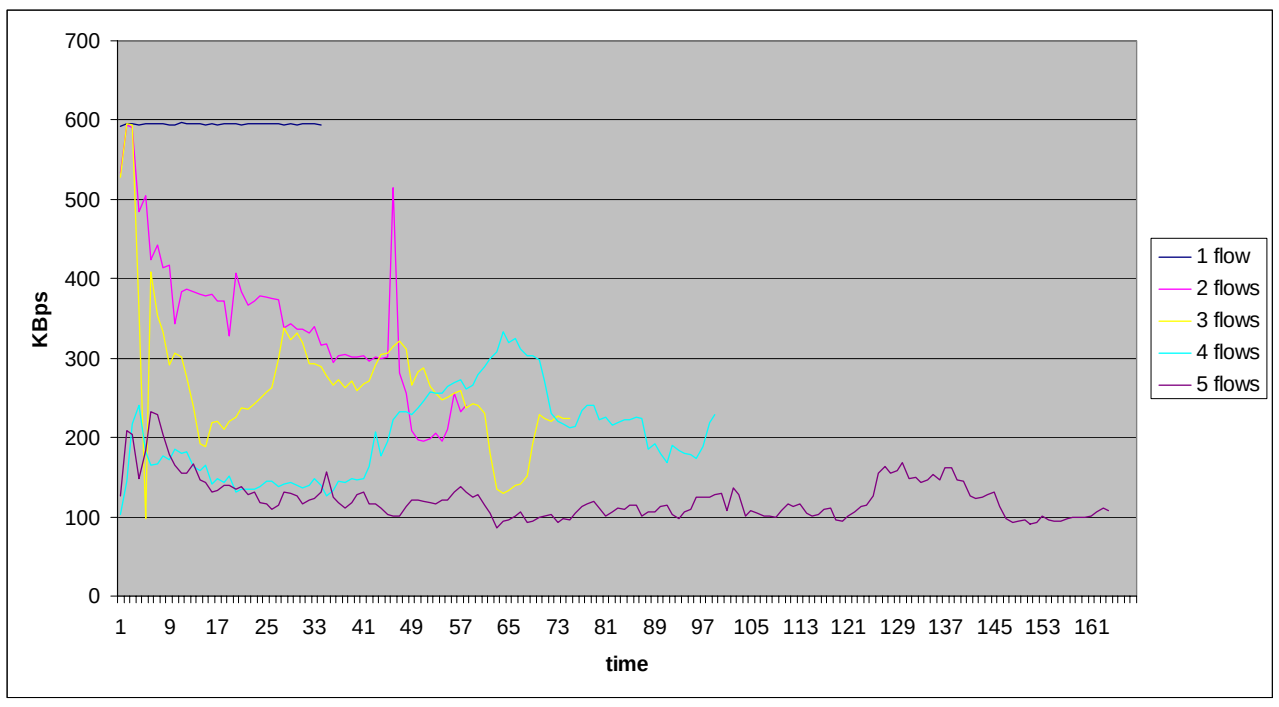

Fig. 3: Throughput (KB) vs. time (s) for transmitting 20000KB with varying number of flows of TCP-SACK with New-AIMD.

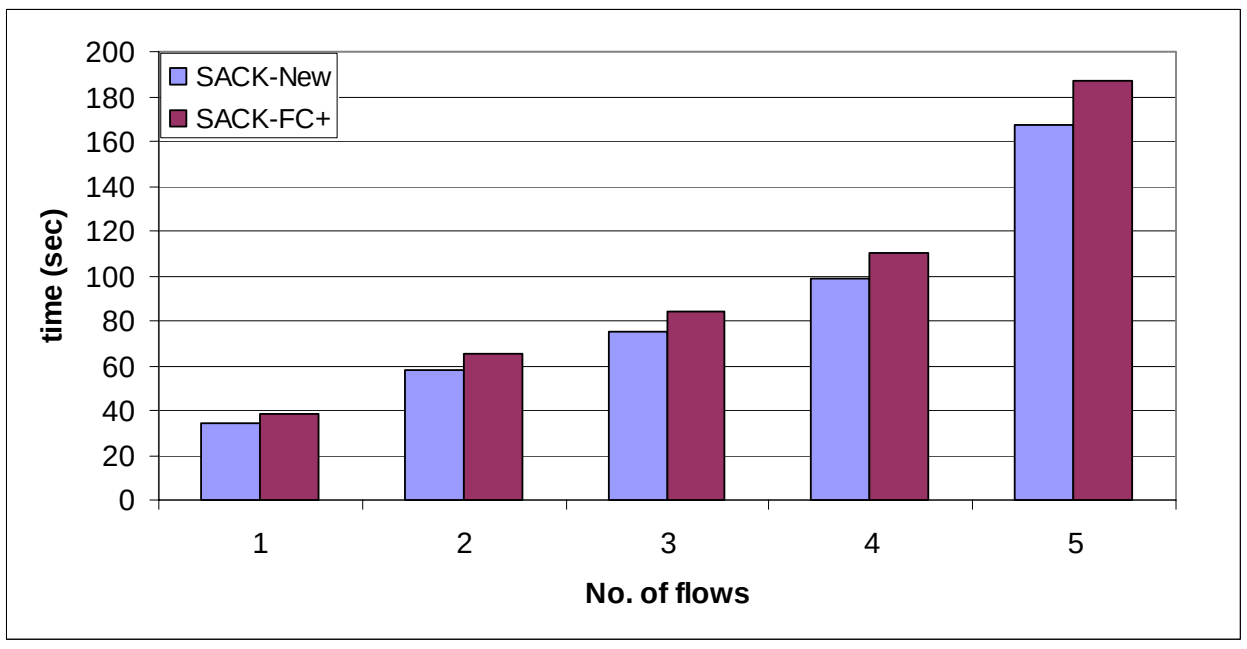

Fig. 4: The time (sec) needed to transmit the 20000KB depending on the number of flows of TCP-SACK with AIMD-FC+ and New-AIMD.

The results for the time needed to transmit the data were less than the expected time needed if we implement AIMD, as in the related work, by around 12% less, which means the end-to-end delay was less as well.

### 8.2 The Second Experiment: Results

In the following figures (fig. 5, 6, 7, 8 with (a, b)), we observe the behavior of the queue using our mechanism (New-AIMD). In this experiment, we measure every RTT. We show that this mechanism works very well, under the given network conditions. And the results will depend on the number of flows, the time for the packet waiting in the queue as the average delay time (ms) and on the queue length in simulation time unit (s). In figures 5, 6, 7 and 8 below we separated the results depend on the number of flows in the experiments (2, 3, 4, or 5 flows) sequentially, and we will put the result in two parts, part a: to show the results for average delay (ms) with the RTT, part b: to show the results for the queue length (bytes) with the simulation time (s).

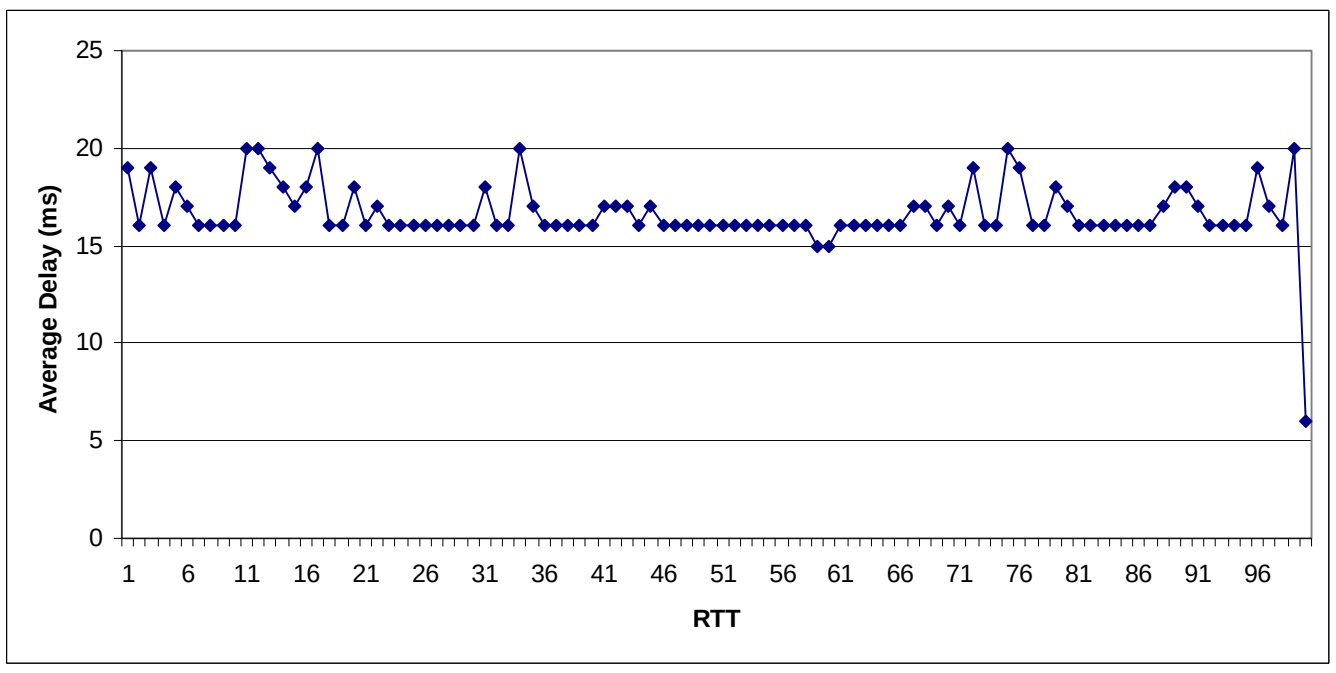

*Fig. 5, a: Average Delays vs. RTT for two flows*

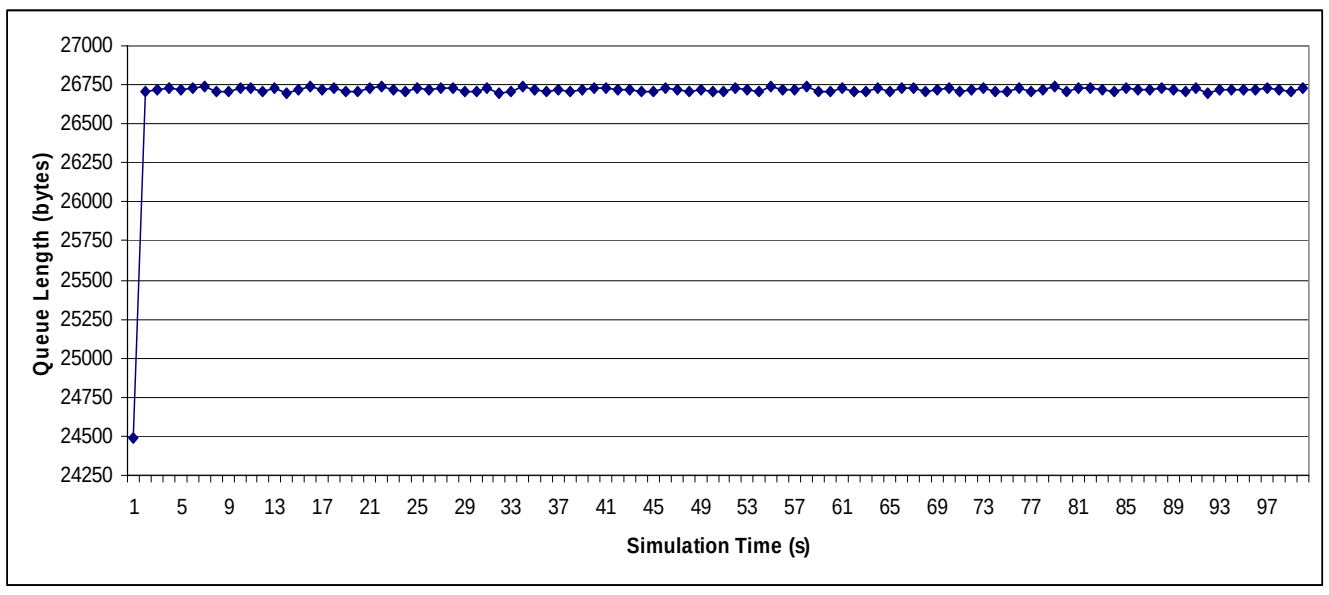

*Fig. 5, b: Queue Length vs. Simulation Time for two flows*

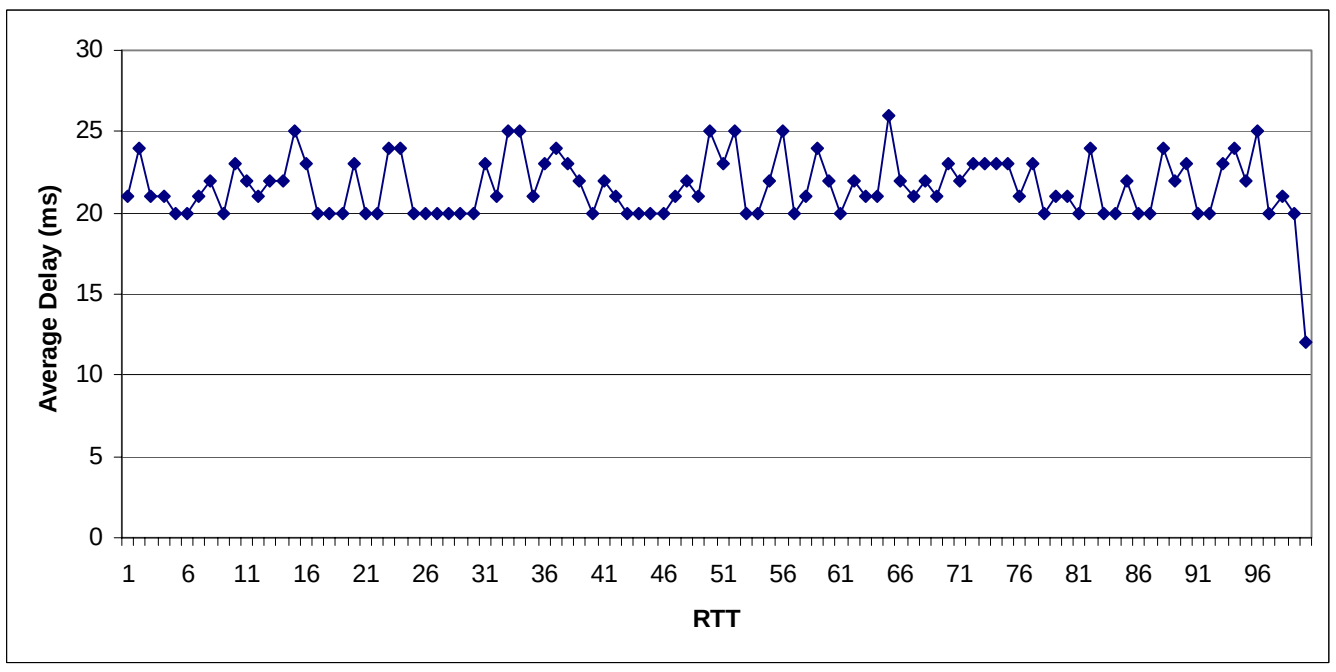

*Fig. 6, a: Average Delays vs. RTT for three flows*

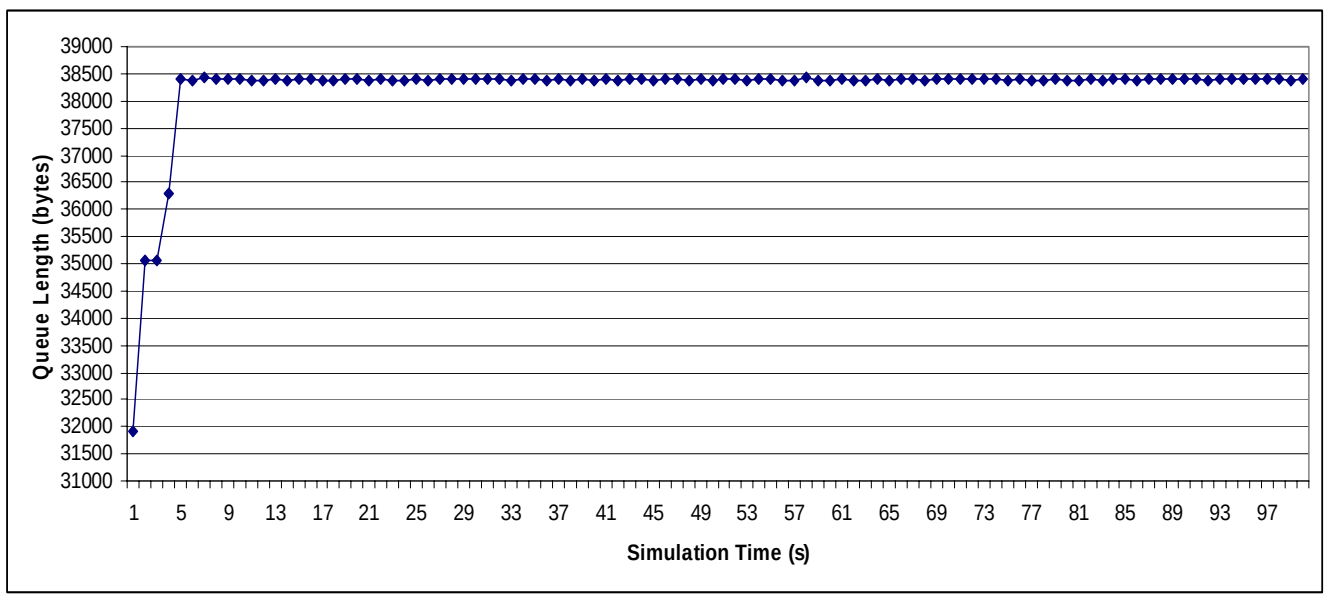

*Fig. 6, b: Queue Length vs. Simulation Time for three flows*

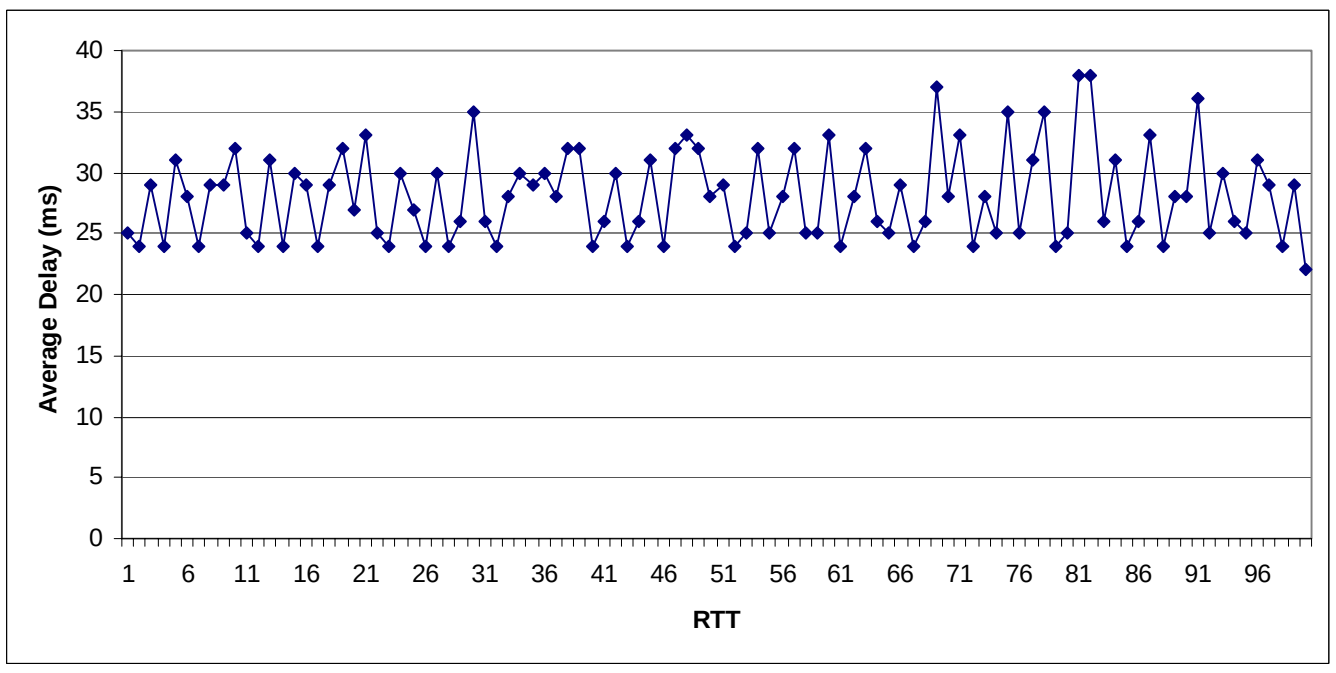

*Fig. 7, a: Average Delays vs. RTT for four flows*

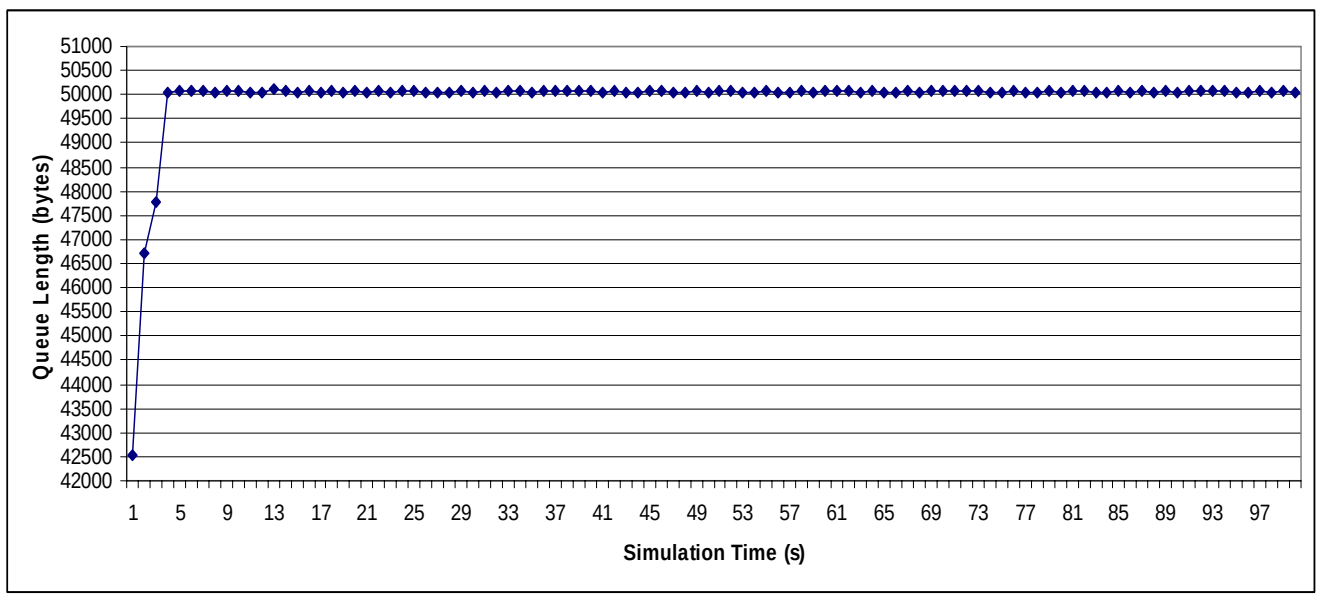

*Fig. 7, b: Queue Length vs. Simulation Time for four flows*

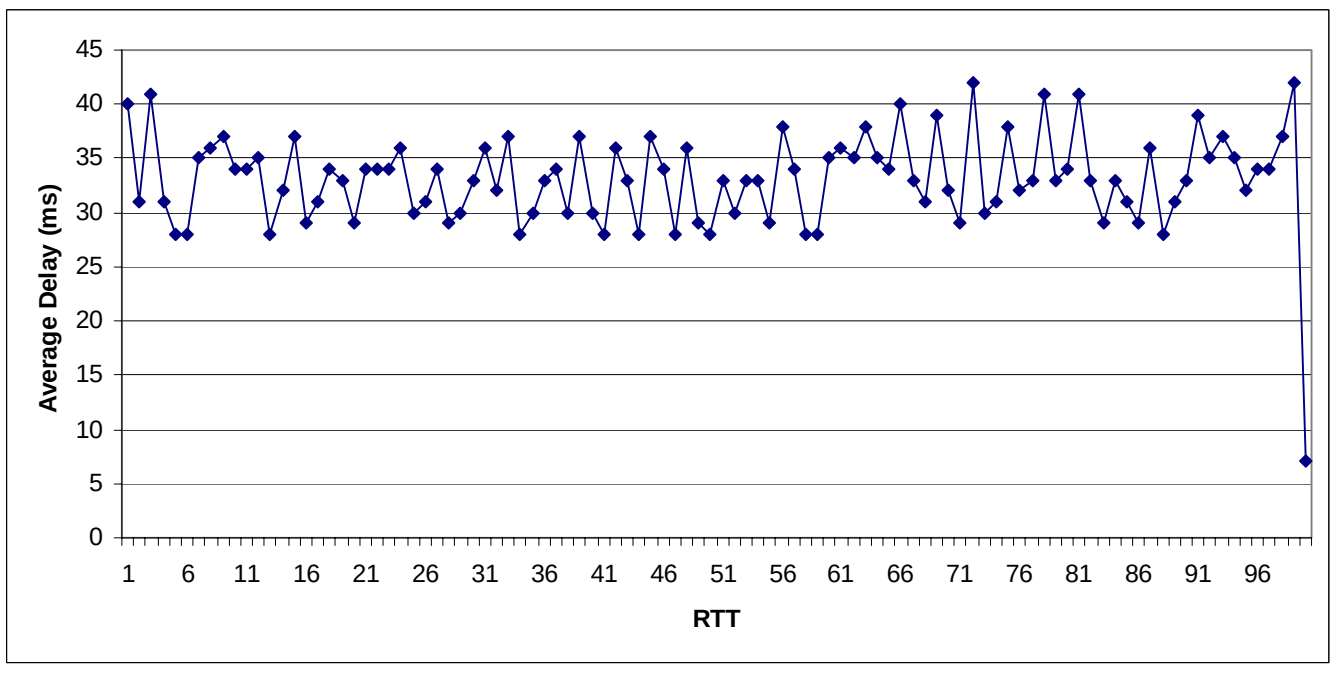

*Fig. 8, a: Average Delays vs. RTT for five flows*

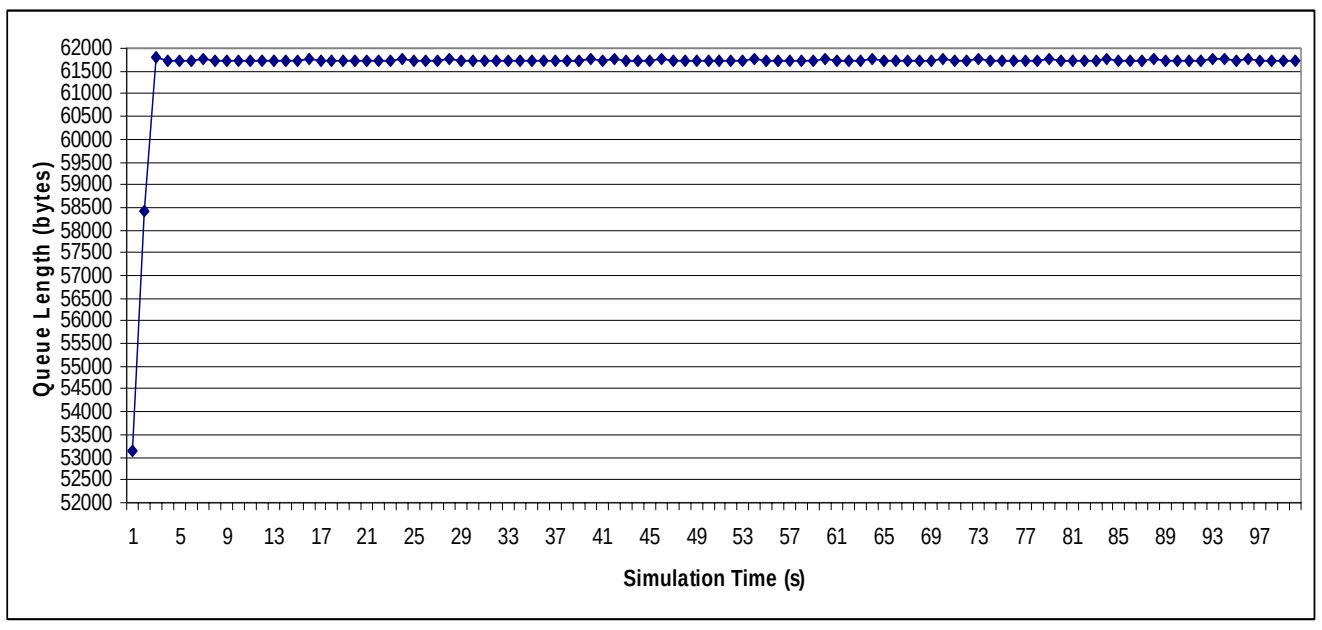

*Fig. 8, b: Queue Length vs. Simulation Time for five flows*

### 8.3 The Third Experiment: Results

In the following figures (fig. 9, 10, 11, 12), we observe the behavior of the using of bottleneck link capacity (utilization) with our mechanism (New-AIMD). In this experiment, we measure the link utilization in every second. We show that this mechanism works very well, under the given network conditions. And the shown results depend on the number of flows. In the table (1) and (fig. 13), we show the average of bottleneck link utilization vs. number of flows use this link in same time, and we measured this average after five second from the start time of the experiment, when the transport rate is stable.

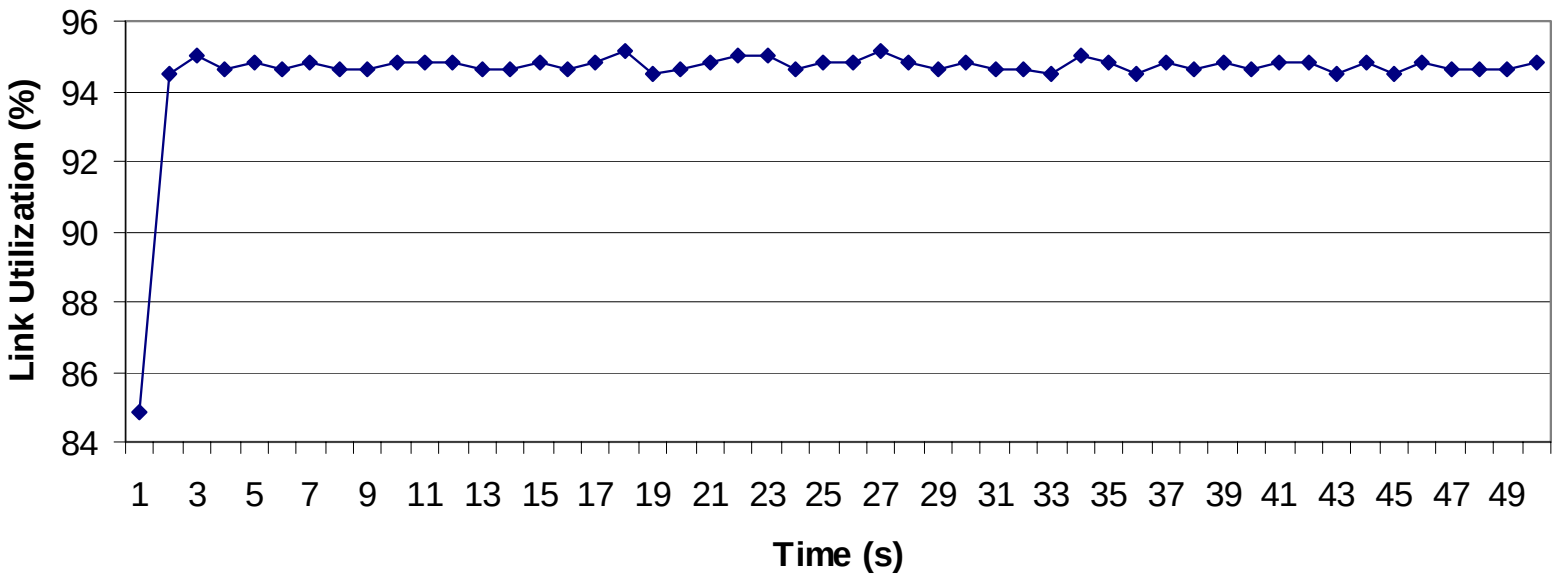

*Fig. 9: Link utilization for bottleneck link with two flows.*

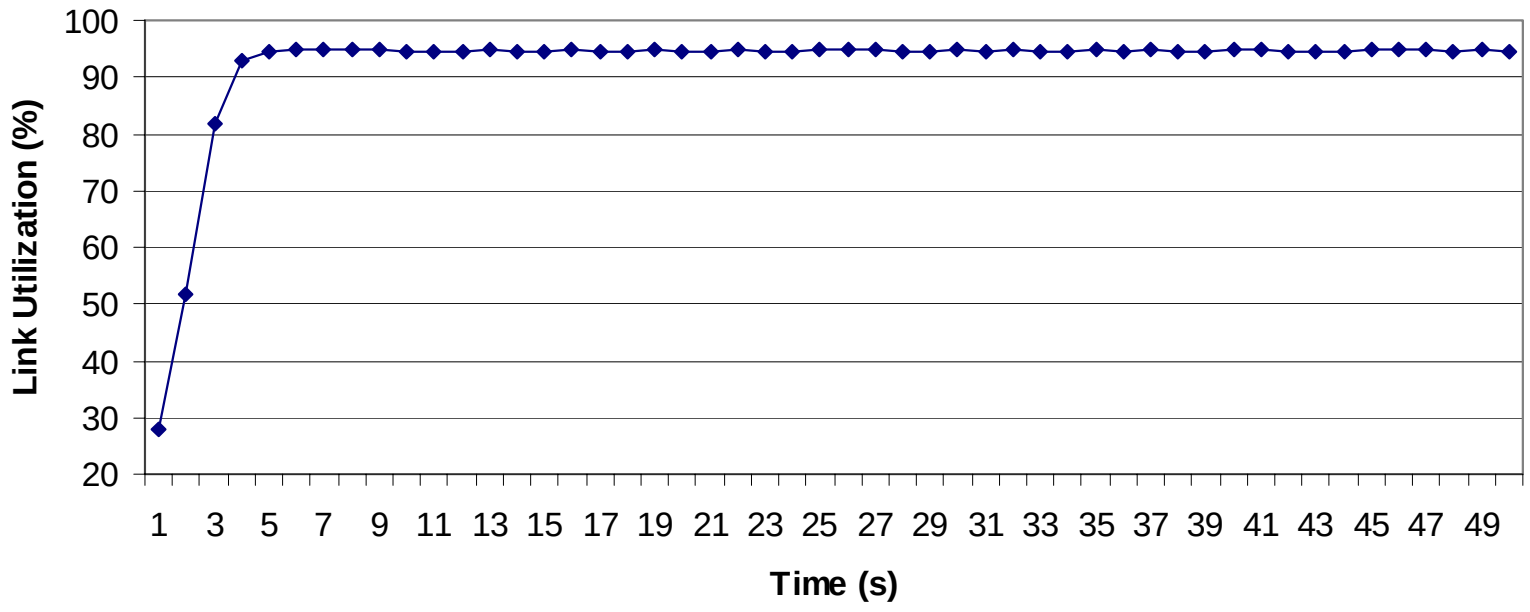

*Fig. 10: Link utilization for bottleneck link with three flows.*

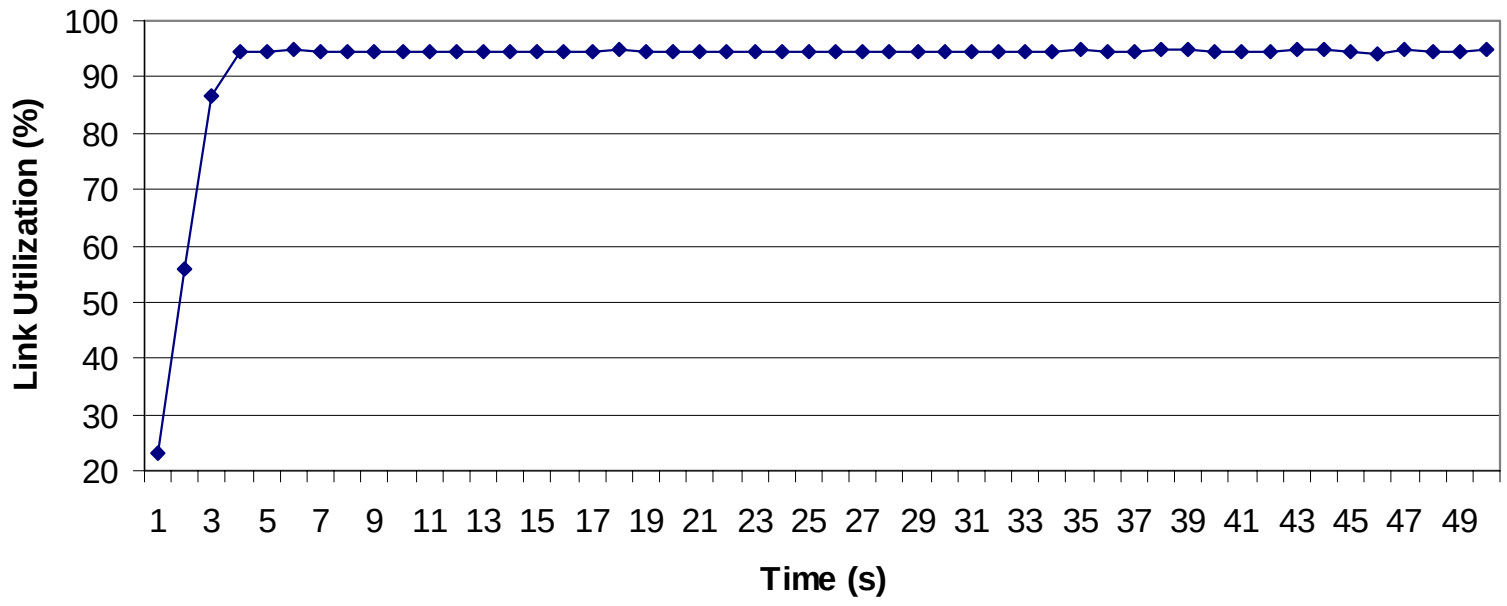

*Fig. 11: Link utilization for bottleneck link with four flows.*

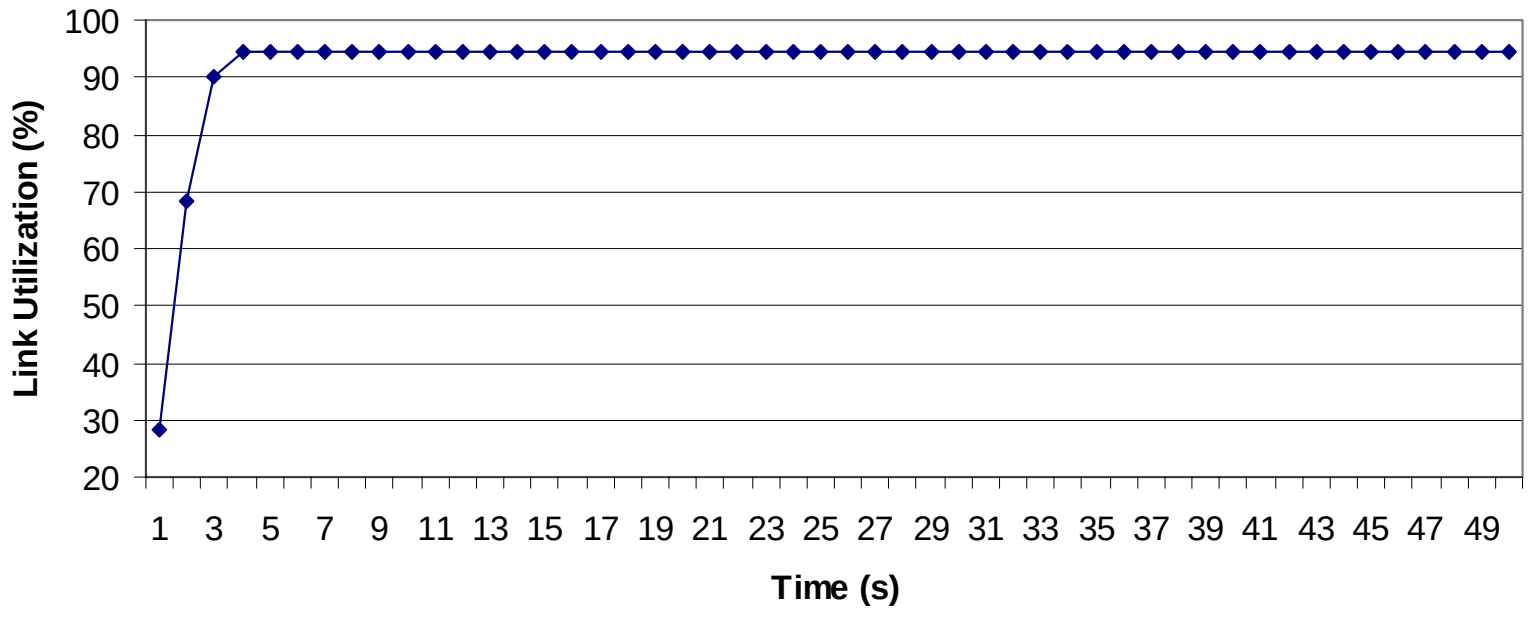

*Fig. 12: Link utilization for bottleneck link with five flows.*

In the figures (fig. 9, 10, 11 and 12) above, we shown the results for the experiments with (2, 3, 4 and 5) flows use the link at the same time, and we can see the average of using the link (utilization) is more than 94%. And we measured this average after five second from the start time; we showed this average in table 1 and figure

13. Also in the figure 13 we shown comparison between the link utilization with our mechanism New-AIMD and AIMD-FC+ in the previous related work [Lahanas, 03].

| No. of flows | Two flows | Three flows | Four flows | Five flows |
|---|---|---|---|---|
| Average of link utilization (%) | 94.73957 | 94.68565 | 94.59935 | 94.48913 |

*Table 1: No. of flows vs. average of bottleneck link utilization with New-AIMD.*

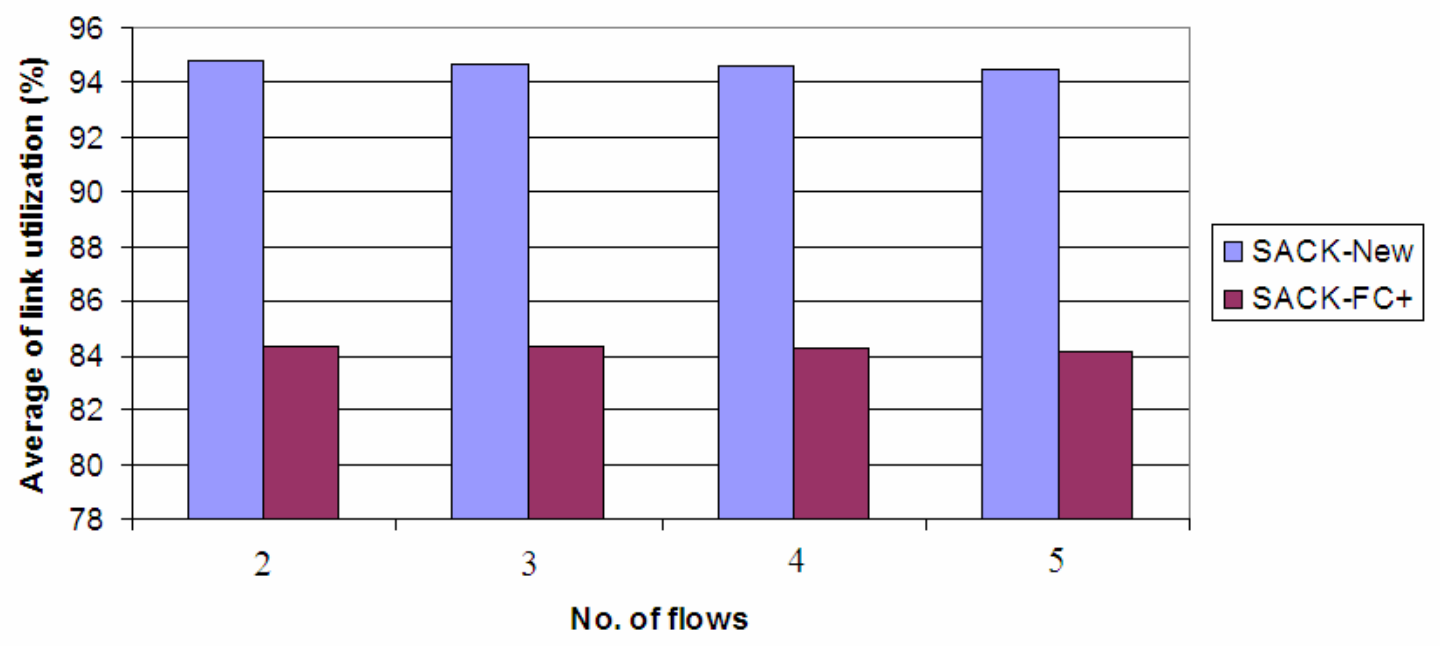

*Fig. 13: Average of link utilization for bottleneck link for different numbers of flows of TCP-SACK with AIMD-FC+ and New-AIMD.*

## 9 Conclusion

In first and second experiments of this work, we investigated about two types of delay, the first one is end-to-end congestion delay; and we found the results after implementing the New-AIMD algorithm were better than the results in the previous work, because the delay was less when we measure the delay depend on the throughput for all the system, and we got end-to-end delay at around 12% less. And we measured the second type of delay, a queuing delay, and also the queue length to discover the bottleneck queue behavior.

And in third experiment, we have experimental evaluation for other performance of New-AIMD mechanism it is the utilization of bottleneck link. We found the results after implement the New-AIMD algorithm and got the high utilization (more than 94%) for the link, and avoid the congestion in this experiment for multi flows use the same link at the same time.

Then we can say this mechanism work as well under the conditions for network experiments above.

And now we can say that the benefit from implementing the New-AIMD algorithm in this study is to reduce the average queue length in order to decrease the end-to-end delay, and also to increase the average of using link bandwidth capacity for network resources (utilization), beside of avoid the network congestion as the major work for this algorithm as we studied it in previous studies about the efficiency and fairness for this mechanism.